# Benchmarking the Processing of Aircraft Tracks with Triples Mode and Self-Scheduling

Andrew Weinert, Marc Brittain, Ngaire Underhill, Christine Serres
MIT Lincoln Laboratory
Lexington, MA 02420
{andrew.weinert, marc.brittain, ngaire.underhill, christine.serres} @ll.mit.edu

*Abstract*—As unmanned aircraft systems (UASs) continue to integrate into the U.S. National Airspace System (NAS), there is a need to quantify the risk of airborne collisions between unmanned and manned aircraft to support regulation and standards development. Developing and certifying collision avoidance systems often rely on the extensive use of Monte Carlo collision risk analysis simulations using probabilistic models of aircraft flight. To train these models, high performance computing resources are required. We've prototyped a high performance computing workflow designed and deployed on the Lincoln Laboratory Supercomputing Center to process billions of observations of aircraft. However, the prototype has various computational and storage bottlenecks that limited rapid or more comprehensive analyses and models. In response, we've developed a novel workflow to take advantage of various job launch and task distribution technologies to improve performance. The workflow was benchmarked using two datasets of observations of aircraft, including a new dataset focused on the environment around aerodromes. Optimizing how the workflow was parallelized drastically reduced the execution time from weeks to days.

## I. INTRODUCTION

The continuing integration of unmanned aircraft system (UAS) operations into the National Airspace System (NAS) requires new or updated regulations, policies, and technologies to maintain safe and efficient use of the airspace. One such technology is detect and avoid (DAA), which enables unmanned aerial systems (UAS) to comply with operating rules and regulations for minimizing the risk of a midair collision (MAC) between aircraft. For operations under the jurisdiction of the Federal Aviation Administration (FAA), relevant rules of Title 14 of the Code of Federal Regulations (14 CFR) include Parts Part 91, §.3, .111, .113(b), .115, .123 and .181(b).

Distribution statement A: approved for public release. This material is based upon work supported by the Federal Aviation Administration under Air Force Contract No. FA8702-15-D-0001. Any opinions, findings, conclusions or recommendations expressed in this material are those of the author(s) and do not necessarily reflect the views of the Federal Aviation Administration. Delivered to the U.S. Government with Unlimited Rights, as defined in DFARS Part 252.227-7013 or 7014 (Feb 2014). Notwithstanding any copyright notice, U.S. Government rights in this work are defined by DFARS 252.227-7013 or DFARS 252.227-7014 as detailed above. Use of this work other than as specifically authorized by the U.S. Government may violate any copyrights that exist in this work. This document is derived from work done for the FAA (and possibly others), it is not the direct product of work done for the FAA. The information provided herein may include content supplied by third parties. Although the data and information contained herein has been produced or processed from sources believed to be reliable, the Federal Aviation Administration makes no warranty, expressed or implied, regarding the accuracy, adequacy, completeness, legality, reliability, or usefulness of any information, conclusions or recommendations provided herein. Distribution of the information contained herein does not constitute an endorsement or warranty of the data or information provided herein by the Federal Aviation Administration or the U.S. Department of Transportation. Neither the Federal Aviation Administration nor the U.S. Department of Transportation shall be held liable for any improper or incorrect use of the information contained herein and assumes no responsibility for anyone's use of the information. The Federal Aviation Administration and U.S. Department of Transportation shall not be liable for any claim for any loss, harm, or other damages arising from access to or use of data information, including without limitation any direct, indirect, incidental, exemplary, special or consequential damages, even if advised of the possibility of such damages. The Federal Aviation Administration shall not be liable for any decision made or action taken, in reliance on the information contained herein.

To satisfy these rules, DAA must meet a set of performance requirements which is often assessed using Monte Carlo safety simulations[1]. A foundational technology to these simulations are statistical encounter models of aircraft behavior based on observed aircraft behavior [2]. For many aviation safety studies, manned aircraft behavior is represented using the MIT Lincoln Laboratory (MIT LL) statistical encounter models. These models are dynamic Bayesian networks trained on observations from a variety of surveillance sources of manned aircraft. Training datasets often include hundreds of thousands of flight hours and necessitate the use of high performance computing (HPC) resources for data processing and model training[2]–[5].

*A. Motivation*

To assess the efficacy of a DAA system, Monte Carlo simulations needs to represent the operational environment where the DAA system is deployed. While various models have been developed for decades, many of these models were not designed to model manned aircraft behavior at low altitudes, where smaller UAS are likely to operate. In response, we previously developed new models for manned aircraft operating from 50 to 5,000 feet above ground level (AGL). and not receiving air traffic control services[3], [6], [7]. These models were designed to assess operations away from airport surfaces or outside of airport traffic patterns. The training data was sourced from the OpenSky Network[6], [8], a community network of ground-based sensors that observe aircraft equipped with Automatic Dependent Surveillance-Broadcast (ADS-B) out. ADS-B was initially developed and standardized to enable aircraft to leverage satellite signals for precise tracking and navigation[9]. However, these new models were not designed to model specific behaviors in the terminal airspace environment, such as when aircraft are taking off, landing, operating in the traffic pattern, or transiting through the airspace. Additionally while the model development workflow leveraged HPC resources[3], there were various computational and storage bottlenecks that hindered additional model development.

*B. Scope*

This work considered how aircraft, equipped with a transponder behaved within 8-10 nautical miles of an airport surface in controlled airspace, within the United States and flying between 50 and 3,000 feet AGL. The scope of this work was informed by the needs of RTCA SC-228, an aviation standards development organization.

*C. Objectives and Contributions*

We focused on two objectives identified by the aviation community: (1) develop datasets of observed aircraft behavior and (2) efficiently processes these datasets to create training data for statistical models of manned aircraft behavior. In response, our primary contributions are the development of the required datasets and a novel workflow to enable efficient, scalable, and rapid processing of the data. The dataset contribution includes the development of a new dataset based on open source observations of aircraft operating in the terminal environment around aerodromes. These contributions and the output of the workflow are the inputs to the model training process, which is out of scope for this paper; please refer to other publications about model training[2], [4], [7]. An enabling technology for these improvements was the triples-mode job launch, a unique job launch mechanism developed at the Lincoln Laboratory Supercomputing Center (LLSC), which provides users with more flexibility to manage memory and threads[10].

## II. STORAGE AND COMPUTE ARCHITECTURE

We first briefly overview the storage and compute infrastructure of the LLSC. The LLSC and its predecessors have been widely used to process aircraft tracks and support aviation research for more than a decade.

*A. Storage and Filesystem*

The LLSC HPC systems have two forms of storage: distributed and central. Distributed storage is comprised of the local storage on each of the compute nodes and this storage is typically used for running database applications. Central storage is implemented using the open-source Lustre parallel file system on a commercial storage array. Lustre provides high performance data access to all the compute nodes, while maintaining the appearance of a single filesystem to the user. The Lustre filesystem[11] is used in most of the largest supercomputers in the world. Specifically, the block size of Lustre is 1MB, thus any file created on the LLSC will take at least 1MB of space.

*B. Compute Infrastructure*

The processing described in this paper was conducted on the LLSC TX-Green HPC system. The system consists of a variety of hardware platforms, but we specifically developed, executed, and evaluated our software using Intel Xeon Phi 64-core nodes (xeon64c), each of which has 64 compute cores in a single processor socket laid out in a mesh configuration[12]. These nodes are allocated 3 GB for each job slot. This architecture has previously been benchmarked for a variety of notional use cases but not benchmarked specifically for an aviation use case.

*C. Triples-Mode Overview*

LLSC users can employ triples-mode, a unique job launch mechanism developed at LLSC which provide users with more flexibility to manage memory and threads[10]. It provides fast resource allocation and job execution by aggregating compute tasks to be executed on the same compute node as a single scheduling task in an array job. It implements explicit process placement and affinity control (EPPAC). When scheduling, the job is assigned exclusive use of each of requested compute nodes; this assigned is referred to as the LLSC exclusive mode. Jobs can be batched or self-scheduled.

Compared to a traditional job launcher, triples-mode can improve performance due to increased flexibility to manage memory and threads. However, from an end-user perspective, triples-mode require more upfront job configuration and planning. Triples-mode is governed by three parameters: (1) number of requested compute nodes; (2) number of processes per node (NPPN); and (3) number of threads per process.

At the time of the benchmarking experiment, LLSC end-users were allocated a default of 4096 xeon64c cores, with each allocated 3 GB memory per slot. As of publication, the allocated default is now 8192 xeon64c cores. LLSC exclusive mode allocates cores by multiple the number of nodes by the slots per node. The slots per node are fixed to 64 for the xeon64c. Thus, due to exclusive mode allocation limits, the maximum number of requested compute nodes was 64 nodes. Also, the LLSC recommended NPPN to be 32 or less and a multiple of 8, due to xeon64c memory constraints. We fixed the number of threads per process, although this can be variable.

The requested compute nodes and allocated compute cores may not be equivalent. Specific to our use case, due to the potential large files, we requested 2 slots per job for a total of 6 GB per slot memory limit, which reduced our maximum potential compute cores to 2048. Due to exclusive mode, we couldn't request additional cores, as 2048 cores with 2 slots per core correspond to the maximum allocation of 4096 cores.

*D. Distribution Rules and Self-Scheduling*

Parallelization was implemented using pMatlab[13] for parallel array programming and LLMapReduce[14], which can allocate tasks to parallel processes via block or cyclic distribution. Previous research[3] used the LLSC default of block distribution which distributes equal sized blocks of consecutive tasks. For example, if there are two processes and four tasks, processes #1 would be allocated tasks 1-2 and processes #2 would be responsible for tasks 3-4. Cyclic distribution allocates tasks in round robin fashion, so in the example the first process would be allocate tasks 1 and 3, while the other process would have tasks 2 and 4.

Tasks are allocated either all upfront as batch or dynamically allocated using a self-scheduling. Previous research[3] solely allocated tasks in batches and we observed significant load imbalances across compute processes. In response, we prototyped a simple self-scheduling approach with one managing process and many worker compute processes. First, the manager sequentially allocates initial tasks to all workers as fast as possible. The manager does not pause when sending the initial messages. Workers receive and complete the initial task and then reports back to the manager. The manager will receive the task completed messages, determine if additional tasks need to be allocated, and then will sequentially send tasks to idle workers. While idle, the workers wait 0.3 seconds prior between checking if another task was sent from the manager, and the manager waits 0.3 seconds prior to checking for more idle workers. The LLSC team recommended the 0.3 second duration. This repeats until all tasks are completed. The rate at which workers complete tasks and become idle will depend on task organization; if many workers are idle, workers may wait minutes prior to receiving their next task. For example, tasks can be randomly organized or by expected size of the task. Task organization specific to our use case is elaborated upon in Sections IV and V.

### III. WORKFLOW AND DATASETS

This section overviews the HPC workflow, datasets used, and illustrates some example processed data.

*A. Processing Workflow*

The high level processing workflow has remained largely unchanged since our previous publication [3]. For the various improvements, please refer to the Git commit messages of the open source repository, Airspace-Encounter-Models/em-processing-opensky, on GitHub.com. Processing steps consisted of (1) parsing and organizing the raw data; (2) archiving organized data; (3) processing and interpolating into track segments.

We first identified unique aircraft by parsing and aggregating various national aircraft registries. All registries specified the registered aircraft's type (e.g. rotorcraft, fixed wing single-engine, etc.); the registration expiration date; and an aircraft's ICAO 24-bit addresses, global unique hex identifier of the transponder equipped on the aircraft. Using the registries, we created a four-tier hierarchical directory structure to organize the data. The top-level directory corresponds to the year, followed by aircraft type, then the number of seats; and the lowest level directory was based on the sorted ICAO 24-bit addresses. This hierarchy ensures that there are no more than 1000 directories per level, as recommended by the LLSC, while organizing the data to easily enable comparative analysis between years or different types of aircraft. The hierarchy was also sufficiently deep and wide to support efficient parallel process I/O operations across the entire structure.

This organization step can create many small files, which can lead to significantly large random I/O patterns for file access when hundreds or thousands of concurrent, parallel processes try to access the small files. This generates massive amounts of networks traffic and is undesirable in a cluster environment. To mitigate this, we create zip archives for each of the bottom directories. In a new parent directory, we replicated the first three tiers of the directory hierarchy from the previous step. Then instead of creating directories based on the ICAO 24-bit addresses, we archive each directory from the previous organization step.

Once archived, the data is processed and interpolated into track segments. Processing includes removing track segments with less than ten observations; calculating the above ground level altitude; identifying airspace class; and estimating dynamic rates (e.g. vertical rate) were calculated. Once processed, track segments are ready to be used to train the statistical encounter models.

*B. Overview of Datasets Curated from the OpenSky Network*

We processed two datasets curated from the OpenSky Network using the described workflow. The first dataset is an extension of the dataset previously used[3] and a new dataset was developed based on location of aerodromes (i.e., airports). We will refer to the first as the Monday dataset and the second as the aerodrome dataset. The datasets differ in temporal and spatial scope, along with the frequency of observations.

The OpenSky Network usually makes easily accessible the global, abstracted, raw state data from the most recent 10-15 Mondays. Each day is organized into 24 files, each corresponding to one UTC hour. Observations are at least ten seconds apart and there is no guarantee on data availability. MIT LL has aggregated most of this data on the LLSC since 2018. A shell script is used to download the data from a public OpenSky Network server. The first dataset consists of 104 Mondays spanning from 2018-02-05 to 2020-11-16. Not all Mondays in this span were included. A previous Monday-based dataset consisted of 85 Mondays across 2018-2019[3]. The Monday-based dataset was originally developed to train models of aircraft not receiving air traffic control services[7].

The second dataset was developed to solely support RTCA SC-228 and to train a model of aircraft operating in the terminal environment of aerodromes. RTCA SC-228 is developing performance requirements for DAA systems to enable UAS to operate in and out of aerodromes[15]. RTCA SC-228 defined the terminal environment as a cylinder with a radius of eight nautical miles (14,816 meters), a height of 3000 feet AGL (914 meters), and centered over an aerodrome. Thus, unlike the first dataset, this aerodrome dataset had specific spatial requirements. To develop this dataset, specific queries to the OpenSky Network Cloudera Impala database were required.

The OpenSky Network offers a historical database using Cloudera Impala and requiring terabytes of storage. Impala is a distributed query engine and does not index structures for query optimization. To improve query efficiency, the data is partitioned in hour-batches by the hour-field of the aircraft observations. Queries can be formulated based on mean sea level (MSL) altitude, time, latitude, longitude, and the ICAO 24-bit address. The raw observations are only in MSL altitude and the OpenSky Network does not estimate the AGL altitude for any observations. However, unlike the Monday state data, observations can be one second, instead of ten seconds, apart. In response, we developed and publicly released software[16] to generate queries based on AGL altitude, location of aerodromes, airspace class, and time zones.

The software[16] generates and organizes queries into three-dimensional bounding boxes. This was the most efficient approach because the OpenSky Network Impala Shell did not support geometric types or functions, such as implemented in PostgreSQL. We could not create queries based on the intersection of polygons and points.

Accordingly, to generate the bounding boxes, the software first identifies all relevant aerodromes and generates a circle with a fixed radius around all aerodromes. Given the RTCA SC-228 scope, we set the radius to eight nautical miles. Next, the circles are combined to create a set of polygons; polygons may overlap and may not be convex. Then bounding boxes are created for each polygon and these boxes are union to create a set of discrete, nonoverlapping, rectilinear polygons, as illustrated by Fig 1.

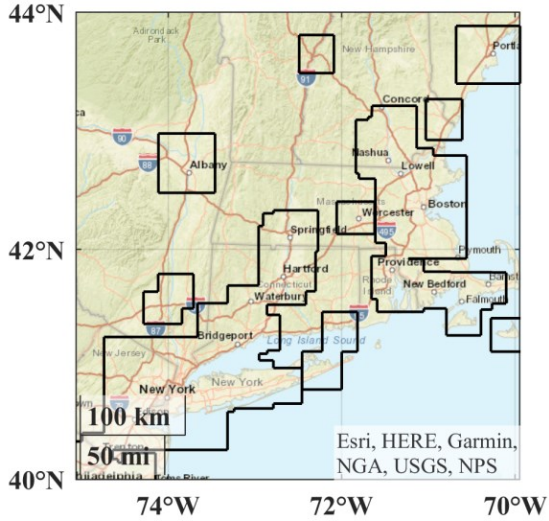

Fig 1. Rectilinear polygons as part of the query generation process.

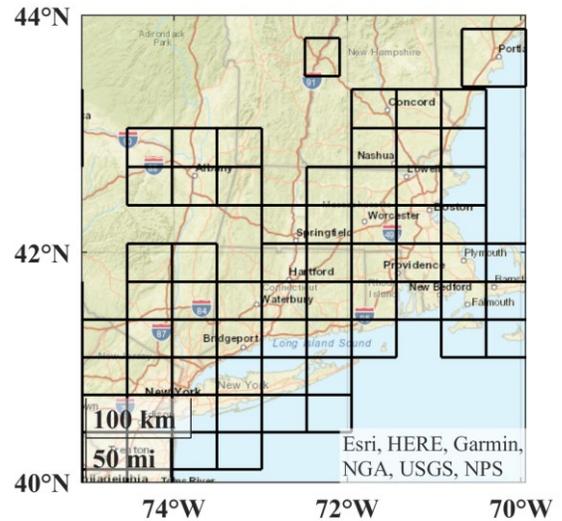

Fig. 2 Example bounding boxes of queries.

Next, these rectilinear polygons are iteratively joined to create simple, nonoverlapping rectangular bounding boxes. For large rectangles, they are iteratively divided into smaller boxes. Each smaller discrete bounding box is then assessed if it is within a desired airspace or distance from aerodrome; bounding boxes that fail these conditions are removed. With bounding boxes established, the NOAA GLOBE[17] digital elevation model is used to estimate the minimum and maximum elevation for each bounding box. The elevation data is used to calculate the MSL range of a query, given a desired AGL altitude range. By default, the desired AGL range is 5,100 feet AGL with a hard ceiling of 12,500 feet AGL. Lastly, the meridian-based time zone is identified for each bounding box and a set of queries are generated based on the bounding box, MSL altitude, and local time. Each query is assigned a group to facilitate load balancing and storage optimization. Fig 2 illustrates the final bounding boxes for the northeastern United States.

Using this software, we generated 136,884 queries for 196 days across 695 bounding boxes across Class B, C, and D airspace across the United States. Temporally, we queried for the first 14 days of each month from January 2019 through February 2020. This time window was largely unaffected by the COVID-19 pandemic, as the Schengen Area travel ban didn't take effect until March 2020[18]. These queries were then executed using a shell script on the MIT Supercloud[10]. Queries were executed in serial across 3-5 CPUs, with a file created for each query. They were recorded in a .txt format and then formatted into .csv. The .csv files were then transferred from MIT SuperCloud to LLSC TX-Green using rsync via a special high bandwidth link. The data was not compressed or archived prior to transferring, with transferring requiring about 5-6 hours. We notionally saw speeds of at least 33 MB/S based on rsync -p output. Benchmarking rsync was out of scope.

*C. Comparson of Datasets*

The datasets differed in spatial and temporal scope, frequency of updates, and storage. Dataset #1, "Mondays," had a global scope with at least 10 seconds between observations, no altitude filtering, and stored across 2425 files organized by day and hour, requiring 714 Gigabytes of storage. Dataset #2, "Aerodromes," was limited to specific regions near USA aerodromes, with observations at least 1 second apart; the altitude observations were filtered on observed MSL and estimated AGL altitudes, and stored across 136,884 files, organized by day and bounding box, requiring 847 Gigabytes of storage. Fig 3. illustrates that the distribution of file sizes also differed between the datasets. Dataset #1 had fewer but larger files; the Gaussian shape was indicative of diurnal pattern due to data organized by hour. Conversely, the sloping distribution of Dataset #2 was indicative that aircraft activity or surveillance coverage is not uniformly distributed across locations; while also introducing load balancing challenges of many small files.

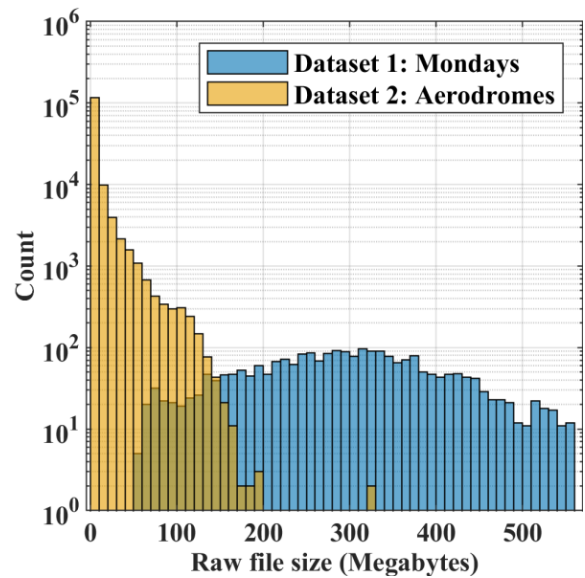

Fig. 3 Distribution of file sizes for each dataset. The bin width is 10 MB.

## IV. PARALLELIZATION AND BENCHMARK RESULTS

Given the established workflow and datasets, this section discusses parallel job optimization, and benchmark results.

### A. Dataset Organization

The first benchmarking experiment was parsing and organizing dataset #1. Job tasks were created for each of the 2425 files. Tasks were allocated via self-scheduling, with one task allocated at a time to workers, and tasks were organized either chronologically or be size. Chronological organization had the earliest date as the first task and the most recent date as the last task. Size organization had the largest file first and the smallest file last. In addition to varying task organization, we also varied the number of requested compute nodes and NPPN. Discussion focuses on dataset #1 but we observed similar benchmarking trends with dataset #2.

TABLES 1 and 2 report the total job time to complete all tasks, as measured by the manager. Foremost, organizing tasks by size always outperformed chronological task organization. When holding the requested compute nodes constant, minimizing NPPN also improved performance. However due to constraints imposed by LLSC exclusive mode, we were only able to request 512 compute nodes when NPPN = 8.

TABLE I. JOB TIME (SECONDS) TO ORGANIZE DATASET #1 WITH CHRONOLOGICAL ORGANIZATION AND SELF-SCHEDULING

| NPPN | *Allocated Compute Cores* | | | |
|---|---|---|---|---|
| | *2048* | *1024* | *512* | *256* |
| 32 | 5640 | 5944 | 7493 | 11944 |
| 16 | - | 5963 | 7157 | 11860 |
| 8 | - | - | 6989 | 11860 |

TABLE II. JOB TIME (SECONDS) TO ORGANIZE DATASET #1 WITH LARGEST FIRST ORGANIZATION AND SELF-SCHEDULING

| NPPN | *Allocated Compute Cores* | | | |
|---|---|---|---|---|
| | *2048* | *1024* | *512* | *256* |
| 32 | 5456 | 5704 | 6608 | 11015 |
| 16 | - | 5568 | 6330 | 10428 |
| 8 | - | - | 6171 | 10428 |

Fig 4. illustrates that requesting more compute cores does not necessarily improve performance and that optimizing NPPN and task organization are important. Notably, 1024 compute nodes with file size organization and NPPN=16 outperformed 2048 compute nodes with chronological organization and NPPN=32. Optimizing NPPN and task organization enabled a 50% reduction in compute nodes while maintaining the same level of performance. These results also illustrate that performance gains due to job optimization are more pronounced when requesting less compute nodes.

Fig. 5-6 report the execution time for each worker, rather than the total job time and when self-scheduling with one manager and 255 workers. They also illustrate that reducing NPPN shifts the distribution to faster times, rather than changing the distribution's shape. Comparing the two figures,

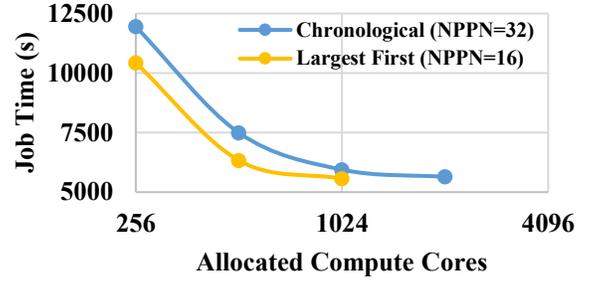

Fig 4. Job time for parsing and organizing dataset #1.

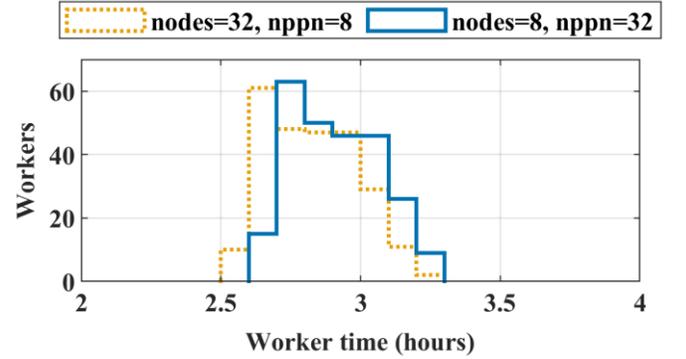

Fig. 5 Distribution of time spent by workers organizing dataset #1 with chronological organization and self-scheduling.

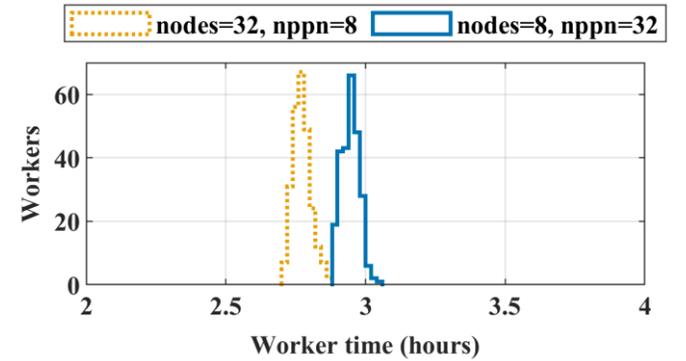

Fig. 6 Distribution of time spent by workers organizing dataset #1 with largest first organization and self-scheduling.

organizing tasks by size reduced the variance of the worker time distribution and minimized the time span between the slowest and fastest workers. In comparisons to the batch task allocation with block distributions of the previous research[3], self-scheduling and triples-mode led to better load balancing and the median worker time decreasing by 14%. The same conclusions are applicable for other triple-modes configurations.

Furthermore, the manager can send multiple tasks per message to workers. We briefly investigated if increasing the tasks per message improved performance. We experimented with organizing dataset #1 using one triples-mode configuration (Nodes = 64, NPPN = 8. Threads = 1) and a cyclic task distribution. The following figure illustrates the results of a performance decrease as tasks per message increase. Since performance degraded, we did not investigate further in this study and identify this as potential future work.

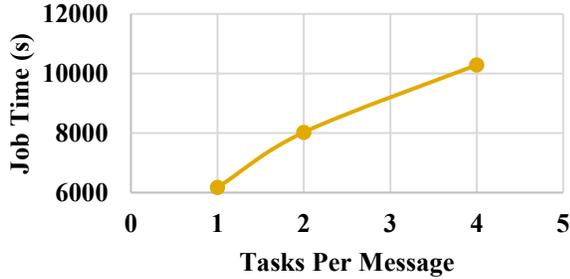

Fig 7. Job time for parsing and organizing dataset #1.

## B. Archiving Organized Data

After the datasets are parsed and organized, they are archived for efficient HPC storage. We previously parallelized this process using a block distribution that resulted in poor load balancing[3]. For the predecessor to dataset #1, 2% of parallel processed account for more than 95% of the total job time, with archiving requiring days to complete. For this effort, we switched to a cyclic distribution which reduced the total job time by more than 90%, enabling archiving to be completed in hours. Since LLMapReduce sorts tasks based on filename, our hierarchical directory structure resulted in tasks effectively sorted by specific aircraft. Tasks associated with aircraft with many observations were sequentially ordered. With block distribution, a worker could be allocated many large tasks for a well observed aircraft while another worker could be allocated all small tasks for aircraft with limited data. By switching to cyclic distribution, this significantly improved load balancing and performance. We observed a similar speed for both datasets, with no notable differences between them.

## C. Processing and Interpolating into Track Segments

Benchmarking the organization and archiving steps consistently indicated that using triples-mode, reducing NPPN and self-scheduling distribution produced better performance. Based on this, we did not iterate over triples-mode parameters. Instead, we benchmarked performance using only 64 requested nodes, NPPN of 16, and a single thread. For self-scheduling, we experiment with randomly organizing tasks instead of by size. Chronological organization wasn't an option because tasks represented specific aircraft rather than spatial or temporal information.

For dataset #2, the median worker time was 13.1 hours, 99.1% of workers finished within 18 hours, 99.7% of workers finished in 24 hours, and all workers completed in 29.6 hours. There was a 17.3 hours difference between the fastest and slowest workers and 16.5 hours between the slowest worker and median. A similar trend was observed for processing dataset #1. However, batch job distribution without self-scheduling or triples-mode required more than 7 days to complete. While load balancing with self-scheduling can be improved, the change to self-scheduling and triples-mode fundamentally improved performance to enable more rapid analyses and exploiting of the data.

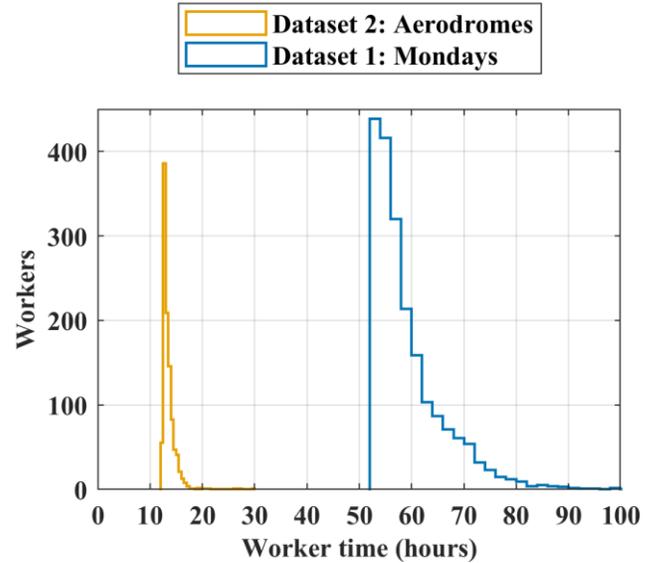

Fig. 8 Distribution of time spent by workers processing the archived datasets with random organization and self-scheduling.

## V. FOLLOW-UP BENCHMARK RESULTS

Based on the benchmarked performance for processing the OpenSky Network-based datasets, a different dataset of aircraft observations was processed to evaluate if lessons learned can be applied across different sources of aviation data. This later benchmarking also benefited from an upgrade to the LLSC that allocated 8192 xeon64c cores to each end-user. Triples mode was configured for 128 nodes, NPPN = 8, two threads, a single 3 GB memory slot for each worker. This configuration was based on the lesson learned of minimizing NPPN but an additional thread was employed based on feedback from the LLSC team.

Observations of manned aircraft were sourced from raw secondary radar reports from terminal radars (ASR-9) that are part of the TCAS RA Monitoring System (TRAMS) over the period January through September 2015. Unlike the OpenSky Network data, these terminal area radar observations are not freely or easily accessible. Specifically, this included the radar located at MIT LL and radars associated with the following airports: KATL, KDEN, KDFW, KFLL, KHPN, KJFK, KLAS, KLAX, KOAK, KORD, KPDX, KPHL, KSDF, KSEA, KSTL. The specific radar identifiers were ATL, DEN, DFW, FLL, HPN, JFK, LAS, LAX, LAXN, MOD, OAK, ORDA, PDX, PHL, PHX, SDF, SEA, STL. The quantity and temporal scope varied across radars. For example, KDFW had data from January through August while KOAK only from June through August. These radar reports provide latitude, longitude and barometric altitude for transponder-equipped aircraft within the radar's surveillance volume. However, this dataset does not include an aircraft's ICAO 24-bit address;

rather the specific ICAO 24-bit address was deidentified and replaced with 13,190,700 generic identifiers. The directory hierarchy remained four tiered but structured based on year, radar location (instead of aircraft type), month range (instead of number of seats), and unique id (instead of ICAO 24-bit address). Additionally, given the scope, an altitude ceiling of 10,000 feet MSL was enforced during data organization.

The lack of ICAO 24-bit address information resulted in significantly more tasks due to organization based on unique id. Suppose a fixed-wing multi-engine flew a round trip (2 flights) between KATL to KDFW. In the OpenSky Network-based dataset, all observations of the 2 flights would be organized in a single task associated with that specific aircraft. However, in this terminal radar-based dataset, the arrival and departure from each airport would be assigned different ids resulting in four tasks. Tasks were randomly ordered and allocated to workers via self-scheduling. Due to the increased quantity of tasks and based on heuristic testing, workers were allocated 300 tasks per self-scheduling message. Thus, there was 43,969 total messages allocated to workers.

Each task consisted of querying a SQL database to organize the data and then processing and interpolating the data into track segments. This was analogous to steps 1 and 3 from the described workflow (Section III.A). New software was developed for the database queries but the same software from em-processing-opensky was used to generate track segments.

This HPC configuration did not have any significant load balancing issues, unlike when previously processing the OpenSky Network-based datasets. The median worker time was 24.34 hours (87,633 seconds) and the span was only 1.12 hours (4057 seconds). The empirical CDF distribution of worker time is provided as Fig. 9. Neither the performance degradation with multiple tasks per self-scheduling message or a significant time span between workers.

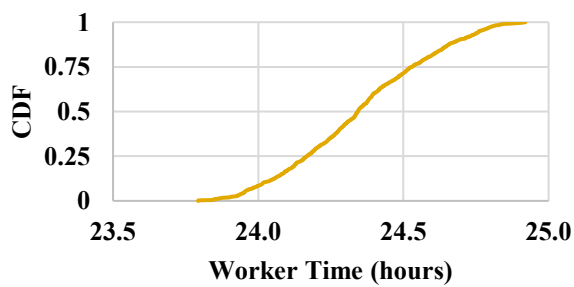

Fig 9. Worker time for organizing and processing radar observations.

This relative performance improvement was primarily attributed to the spatial and temporal scope for each task. Since each task was limited to continuous observations of a given aircraft surveilled by one sensor, the computational overhead for each task was relatively smaller than the OpenSky Network-based dataset. Specifically, when calculating the AGL altitude, the amount of DEM data required was constrained by the surveillance range of the radar. Compared to the OpenSky-based tracks that could span hundreds of nautical miles and multiple USA states, potentially requiring significantly more DEM data to be loaded into memory and manipulated. Also, since there are no temporal constraints, the span of observations for each unique id is also more constraint. The frequency in which a specific aircraft operates in a region is less impactful.

More importantly, this benchmark indicates that processing observations of aircraft using a triples-mode configuration with self-scheduling task allocation can be achieved in reasonable timeframe with good load balancing. These results also indicate that the described workflow is appropriate for different sources of aircraft data. Future work for the OpenSky Network-based datasets should focus on improving task generation by leveraging spatial and temporal information.

## VI. CONCLUSION

We substantially improved the performance of a novel workflow to organize, archive, and process observed aircraft tracks. Parallelization optimizations of self-scheduling and triples-mode operation has drastically reduced the execution time from weeks to days. The performance improvements enabled more resources to be focused on downstream efforts. Additional benchmarking is possible future work, as we did not vary the number of threads for the presented research.

While triples-mode is a unique LLSC capability, the lessons learned about organizing and distribution parallel tasks for aircraft track processing are generally applicable. Additionally the time required to fully process the various datasets is also generally applicable. The paper demonstrates that without HPC resources, exeucting the end-to-end workflow on a few cores would require potential thousands of days and would be impracticable. Lastly, many of the capabilities described in this paper have been, or are in the process of being, transitioned as open source software under permissive open source licenses. On GitHub.com, please refer to the MIT Lincoln Laboratory (@mit-ll) and Airspace Encounter Models (@Airspace-Encounter-Models) organizations.

ACKNOWLEDGMENT

This research supported the FAA UAS research task: A11L.UAS.2. We greatly appreciate the support and assistance provided by Sabrina Saunders-Hodge, Deepak Chauhan, and Alex Fu from the Federal Aviation Administration. We also would like to thank fellow colleagues Dr. Rodney Cole and Wes Olson. The authors acknowledge the MIT SuperCloud and Lincoln Laboratory Supercomputing Center for providing HPC and consultation resources that have contributed to the research results reported within this paper.